\documentclass[doublecol]{epl2} 
\usepackage{graphicx}
\usepackage{lipsum}
\usepackage{float}
\usepackage[utf8]{inputenc}
\usepackage{color}
\usepackage{dcolumn}
\usepackage{bm}

\usepackage{amsmath}

\title{Active Colloids in Harmonic Optical Potentials}
\shorttitle{Title} 

\author{I. Buttinoni\inst{1}, L. Caprini\inst{2}, L. Alvarez\inst{3}, F. J. Schwarzendahl\inst{2} \and H. Löwen \inst{2}}
\shortauthor{F. Author \etal}

\institute{                    
  \inst{1} Institute for Experimental Physics of Condensed Matter - Heinrich-Heine-Universität Düsseldorf, Universitätsstr. 1, 40225 Düsseldorf, Germany.\\
  \inst{2} Institute for Theoretical Physics II - Heinrich-Heine-Universität Düsseldorf, Universitätsstr. 1, 40225 Düsseldorf, Germany.\\
  \inst{3} Centre de Recherche Paul Pascal - University of Bordeaux, 115 Avenue du Dr Albert Schweitzer, 33600 Pessac, France.
}

\abstract{We study the motion of active patchy colloids in an optical trap using experiments, theory and numerical simulations. To achieve isotropic and harmonic confinement, we prototype microparticles with a nearly uniform refractive index and verify that, in the absence of activity, the confined motion is identical to that of optically homogeneous Brownian particles. If the activity is turned on by means of vertical AC fields, the density distributions are described by Boltzmann-like statistics (Gaussian with effective temperature) only for strongly confining traps, whereas weaker potentials give rise to non-Gaussian distributions with a bimodal shape. Our results showcase a simple way to study active soft matter in optical potential landscapes eliminating the optical torque.}

\begin{document}

\maketitle

Optical tweezers, {\sl i.e.} tightly focused laser beams, are widely used in soft condensed matter to confine and manipulate micro- and nanospheres that have a refractive index larger than the one of the surrounding medium \cite{Grier_Review,Optical Tweezers,Volpe_Review}. Their success is partly due to the fact that, in a very good approximation, the sum of the forces exerted by the trap in the focal plane normal to the beam's propagation gives rise to a harmonic potential well. Because the confining forces behave as springs of stiffness $k$, the equilibrium spatial probability distribution of a trapped particle is a Gaussian curve centred around the midpoint, and any external force $\rm \textbf{F}$ can be quantified from the average displacement of the particle from the trap's centre. This simple behaviour prompted a plethora of fundamental and applied studies including the measurement of interaction forces with sub-piconewton accuracy (photonic forces microscopic) \cite{Crocker_Entropic,Buttinoni_Mechanical,Park_Heterogeous,Mittal_Polarization,ElMasri_Measuring,Fadanelli_Measurement}, the investigation of resonance and stochastic phenomena \cite{Ciliberto_Experiments,Franosch_Resonance,Williams_Direct,Simon_Escape,Gammaitoni_Stochastic}, and the development of microscopic heat engines \cite{Bickle_Realization,Martinez_Brownian} and information machines \cite{Admon_Experimental}. 

The harmonic approximation is valid so long as the trapped object is homogenous in shape and composition \cite{Optical Tweezers,Simpson_Inhomogeneous} -- a condition that does not usually apply to microswimmers since symmetry breaking is a key ingredient to achieving locomotion at the micro- and nanoscale \cite{Purcell}. Synthetic active colloids, {\sl i.e.} microparticles that undergo self-propulsion rather than being in thermal equilibrium with the fluid \cite{Bechinger_Active}, are often fabricated by depositing onto one hemisphere of bare polymer or silica particles a second material (e.g.\ a layer of metal). The `cap' interacts with the surrounding liquid to generate local slip flows, and thus self-propulsion \cite{Aubret_Eppur}, but at the same time introduces refractive-index mismatches so that, in combination with optical tweezers, the particles can no longer be assumed as confined in a spring-like potential \cite{Aubret_Metamachines,Moyses_Trochoidal,Lawson_Optical,Zong_Optically}. Because of this, the state-of-the-art of light fields applied to active colloids is, so far, mostly limited to optical landscapes that reorient the particles or tune their velocities \cite{Pince_Disorder,Lozano_Phototaxis,Lavergne_Group,Jahanshahi_Realization}, but do not provide a confining harmonic potential (except for the experiments in Ref. \cite{Schmidt_Non-equilibrium}, done with nanoparticles).\ In contrast, the latter situation has been studied theoretically \cite{Caprini_Parental,Szamel2014,Pototsky2012,Jahanshahi_Brownian,Chaudhuri_Active,Khadem_Delayed}, in granular active systems \cite{Dauchot_Dynamics} and in acoustic confinements \cite{Takatori2016Acoustic}.

In this work, we fabricate patchy colloidal particles with a nearly homogeneous refractive index, confine them in a tightly focused laser beam, and activate them by means of AC electric fields applied in the direction perpendicular to the plane of motion. We demonstrate that, despite the patchiness of the particles, the optical trap is harmonic. Within a range of swimming velocities and laser intensities, the active motion in the optical tweezer agrees well with theoretical predictions and numerical simulations of active Brownian particles in harmonic wells, showing non-Boltzmann properties.

The remainder of the manuscript is structured as follows. As a preliminary study, firstly, we consider the active motion of optically homogenous -- yet patchy -- particles without confinements (swimming velocity $\rm V_0 > 0$, trap's stiffness $k = 0$). Secondly, we introduce and characterise the confinement, considering optically-trapped particles that are `passive', {\sl i.e.} there without activity ($\rm V_0 = 0$, $k > 0$). Finally, we focus on active Brownian particles in optical traps ($\rm V_0 > 0$, $k > 0$), interpreting the main result in a final discussion.

\section{Unconfined Active Colloids ($\rm V_0 > 0$, $k = 0$)}

\begin{figure}[t]
\center
	\includegraphics[scale=0.5]{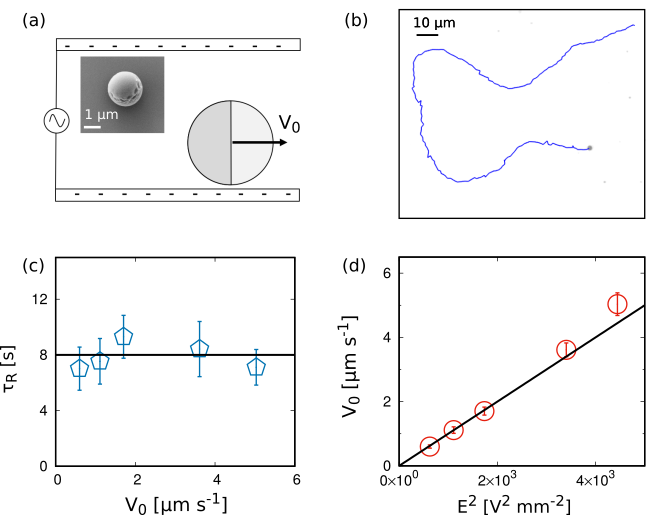}
	\caption{Unconfined active colloids. (a) Schematic illustration of self-propulsion due to induced-charge electroosmotic flows (ICEO) under AC electric fields applied in the vertical direction. Inset: scanning electron microscopy image of a $\rm PMMA$-$\rm SiO_2$ colloidal particle. (b) Two-dimensional trajectory of a $\rm PMMA$-$\rm SiO_2$ particle swimming just above the bottom electrode ($\rm E = 66$ $\rm V/mm$). (c) Characteristic reorientation time, $\tau_{\rm R}$, in the 2D plane of motion as a function of the swimming velocity, $\rm V_0$. (d) $\rm V_0$ plotted against the electric field squared. The line is a linear fit. In (c) and (d) each data point is an average of $15-20$ particles.}
	\label{Fig1}
\end{figure} 

We prepare patchy colloids by coating one hemisphere of polymethyl methacrylate ($\rm PMMA$) microparticles (radius $\rm R \simeq 1$ $\rm \mu m$, refractive index $\rm n_{PMMA}=1.48$, Microparticles GmbH) with a thin layer of silicon dioxide ($\rm SiO_2$, thickness $\lambda = 10$ $\rm nm$, $\rm n_{SiO_2}=1.46 - 1.48$ \cite{Rodriguez_Self,Gao_Exploitation}) using a magnetron sputter coater (PVD Inc.). A scanning electron microscope (SEM) snapshot of the resulting patchy particle is shown in the inset of Fig.~\ref{Fig1}(a) (note that the `wrinkles' are due to partial buckling of PMMA in high vacuum, {\sl i.e.} during $\rm SiO_2$ deposition). The $\rm PMMA$-$\rm SiO_2$ colloidal particles are dispersed in deionised water and pipetted in a thin sample cell (thickness $120$ $\rm \mu m$) made of two planar negatively charged electrodes (indium tin oxide coated glasses functionalised with poly-sodium-4-styrenesulfonate, surface resistivity $8-12$ $\rm \Omega/sq$, Merck KGaA). We apply vertical AC electric fields at frequency $\rm f = 4$ $\rm kHz$ and look at the two-dimensional motion of the particles onto the bottom electrode by means of an inverted light microscope (Olympus IX73).

The $\rm PMMA$-$\rm SiO_2$ microspheres are active because of induced-charge electro-osmosis (ICEO, see the sketch in Fig.~\ref{Fig1}(a)). The electric field, $\rm \textbf{E}$, applied in the vertical direction sets in motion the electrical double layer of the substrate, polarises the particle and, ultimately, produces recirculating hydrodynamic flows \cite{Ristenpart_Electrohydrodynamic}.\ If the particle is patchy, the magnitude (or even the direction) of the rolls near the two faces is different, causing net motion of the particles in the plane parallel to the electrode. This type of self-propulsion was demonstrated for both hybrid dimers \cite{Ma_Inducing,Ni_Hybrid,Alvarez_Reconfigurable} and Janus colloids \cite{Fernandez_Feedback} swimming onto conductive surfaces; remarkably, it does not require a metallic coating, in contrast to other self-propulsion mechanisms such as induced-charge electrophoresis (ICEP) \cite{Gangwal_Induced}.

Figure~\ref{Fig1}(b) shows a typical active trajectory; the $\rm PMMA$-$\rm SiO_2$ particle swims at constant velocity $\rm V_0$ along the axis linking the poles of the two hemispheres and reorients according to its rotational diffusion time $\rm \tau_R= \rm D_R^{-1}=(8 \pi \eta R^3)/(k_B T)$ (Fig.~\ref{Fig1}(c), horizontal line), where $\eta$ is the water viscosity and $\rm k_B T$ is the thermal energy at room temperature. $\rm \tau_R$ and $\rm V_0$ are extracted by fitting the mean squared displacement (MSD) \cite{Bechinger_Active}; $\rm \tau_R$ does not depend on the activity, $\rm V_0$ (Fig.~\ref{Fig1}(c)), whereas $\rm V_0$ increases linearly with the electric field squared, $\rm E^2$ (Fig.~\ref{Fig1}(d)) \cite{Ma_Inducing,Ni_Hybrid}. Due to the lack of optical contrast between the two hemispheres, which is paramount for optical tweezing, we are unable to resolve the local orientation. Nonetheless we speculate that the particles swim with the $\rm SiO_2$ hemisphere heading, in analogy with polystyrene-silica dumbbells \cite{Ni_Hybrid}.

\section{Confined Passive Colloids ($\rm V_0 = 0$, $k > 0$)}

\begin{figure}
	\center
	\includegraphics[scale=0.5]{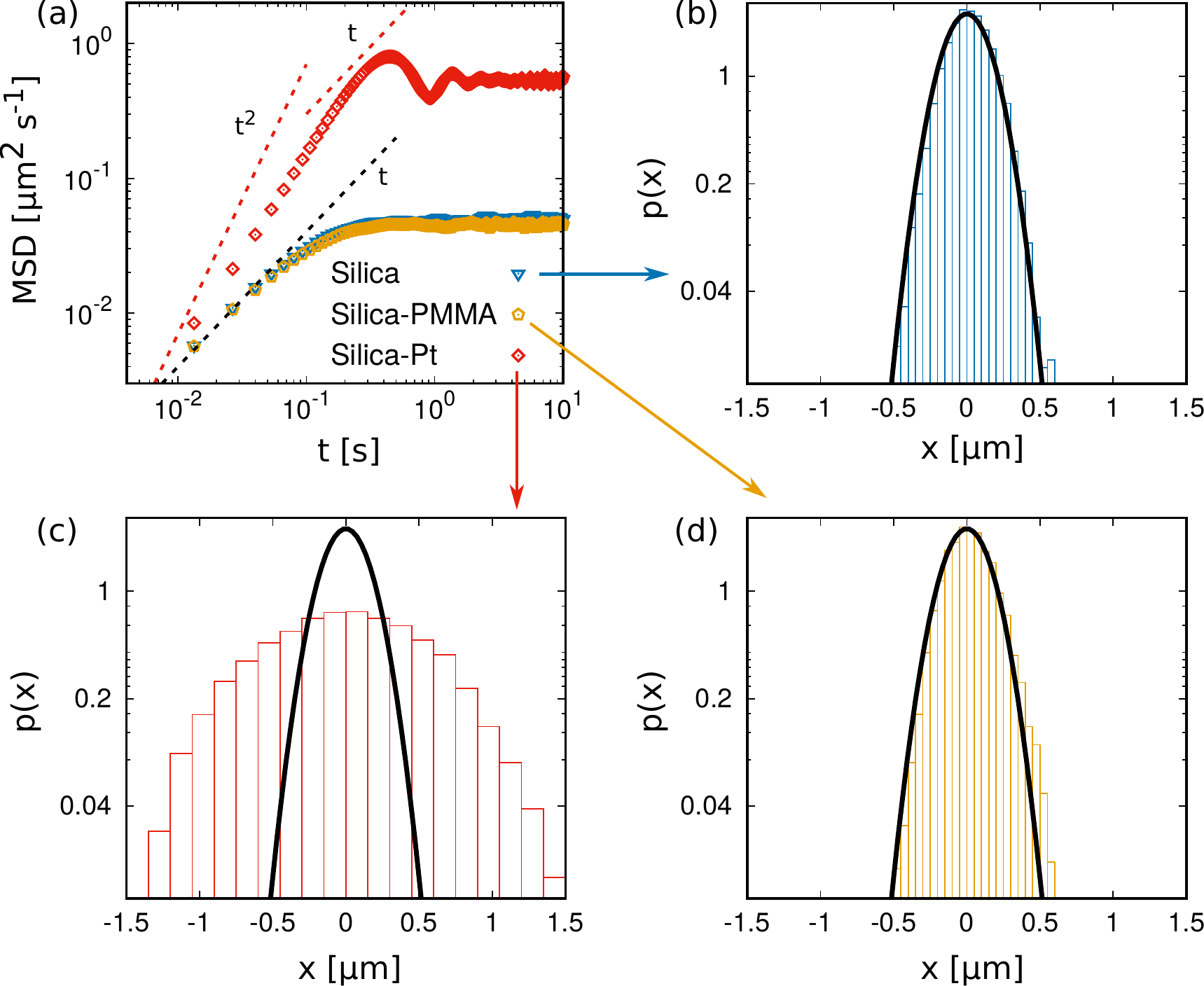}
	\caption{Confined passive colloids. (a) MSD of colloidal particles (see legend) in an optical trap of stiffness $k = 0.18$ $\rm pN/\mu m$. (b-d) Corresponding density distributions evaluated for approximately $\rm N = 10^4$ coordinates. The histograms are normalized: the number of elements in the bin is divided by the number of elements in the input data. The solid black lines represent the Gaussian distributions expected in a harmonic confinement of curvature $k$, as described in the main text.}
	\label{Fig2}
\end{figure}

We now consider the scenario in which the patchy particles are not active ($\rm E=0$, $\rm V_0=0$) and confine them in an optical trap (wavelength, $\lambda = 532$ $\rm nm$). We measure their MSD (Figure~\ref{Fig2}(a), orange curve) and spatial probability distribution ($\rm p(x)$, Figure~\ref{Fig2}(d)), and compare them to those of homogeneous (see blue data in Fig.~\ref{Fig2}(a) and Fig.~\ref{Fig2}(b)) and $\rm Pt$-coated (see red data in Fig.~\ref{Fig2}(a) and Fig.~\ref{Fig2}(c)) $\rm SiO_2$ colloids of similar size. The latter particle is chosen as model {\sl common} synthetic microswimmer since the majority of active microspheres available to date are equipped with metallic caps. Notwithstanding, note that in this Section none of the particles are active.

The dynamics of an optically-homogeneous microsphere in an optical trap centred on the origin is described by a simple overdamped (diffusive) equation of motion for the particle position $\mathbf{x}$:

\begin{equation}
\label{eq:passivedynamics}
\gamma\mathbf{\dot{x}}=\mathbf{F}(\mathbf{x}) + \gamma\sqrt{2 \rm D}\rm \textbf{W},
\end{equation}

\noindent where $\rm \textbf{W}$ is $\delta$-correlated white noise with unit variance and zero average. The constants $\gamma = 6 \pi \eta \rm R$ and $\rm D =k_B T/\gamma$ are the friction and diffusion coefficients, respectively. The term $\mathbf{F}(\mathbf{x})$ describes the confinement force due to the optical trap; in first approximation, it can be recast onto a linear force,

\begin{equation}
\label{eq:forceshape}
\mathbf{F}(\mathbf{x})\approx- k \mathbf{x},
\end{equation}

\noindent where $k$ is the trap's stiffness (or spring constant) which determines the curvature of the harmonic potential. As a result, the density distribution projected onto the $\rm x$-axis, $\rm p(x)$, is described by a Gaussian curve of variance $\sigma^2=\left (\rm k_B T \right)/k$ (black lines in Figs.~\ref{Fig2}(b-d)) and the MSD for the $\rm x$-coordinates has the following form:

\begin{equation}
	\label{eq:MSD}
	{\rm MSD(t)} = 2 \frac{\rm k_B T}{k} \left(1-e^{-\frac{k}{\gamma}\rm t}\right).
\end{equation} 

\noindent It is diffusive for $\rm t \rightarrow 0$ and reaches a plateau at $2{\rm k_B T}/k$ when ${\rm t} \gg \gamma/k$, as usual for passive Brownian particles. Note, however, that the optical trap has a finite size and the harmonic approximation breaks down at the boundary where the shape of potential bends to a maximum. 

The dynamics described above is observed for both bare $\rm SiO_2$ colloids (blue data in Fig.~\ref{Fig2} and Supplementary Video S1) and $\rm PMMA$-$\rm SiO_2$ Janus microspheres (orange data in Fig.~\ref{Fig2} and Supplementary Video S2). In contrast, microparticles with metallic caps behave differently (Supplementary Video S3): the red data in Fig.~\ref{Fig2} indicate that at short timescales (up to $\rm t \approx 1$ $\rm s$) the motion of $\rm Pt$-$\rm SiO_2$ colloids in the trap is superdiffusive, possibly due to additional optical `kicks' as well as self-thermophoretic effects similar to those observed in Ref.~\cite{Moyses_Trochoidal}. As such, the corresponding Gaussian density distribution (Figure~\ref{Fig2}(c)) has a width that is remarkably larger than what is expected in a harmonic confinement of curvature $k$, consequently departing from a Boltzmann distribution (even if Gaussian). Before saturation, the MSD of $\rm Pt$-$\rm SiO_2$ particles also exhibits a clear oscillatory behaviour (Fig.~\ref{Fig2}, red data for $\rm t > 1$ $\rm s$). These oscillations confirm the presence of optical torques, stemming from the fact that the optical gradient force acts more on the $\rm Pt$ cap than the uncoated hemisphere \cite{Aubret_Metamachines,Moyses_Trochoidal,Lawson_Optical,Zong_Optically}. 
	
A comprehensive study of the motion of optically-inhomogeneous particles in laser traps is beyond the scope of this manuscript because it involves effects we want to avoid. Nonetheless, we expect that the behaviour shown by the red data in Fig.~\ref{Fig2} affects all active systems where self-propulsion is achieved by breaking the symmetric optical composition of the particles. This is not the case of $\rm PMMA$-$\rm SiO_2$ colloids, which are optically identical to their homogeneous counterparts; they are therefore a suitable choice to study active motion in combination with harmonic optical potentials. 

\section{Confined Active Colloids ($\rm V_0 > 0$, $k > 0$)}

\begin{figure*}[t]
\center
\includegraphics[width=0.95\textwidth, keepaspectratio]{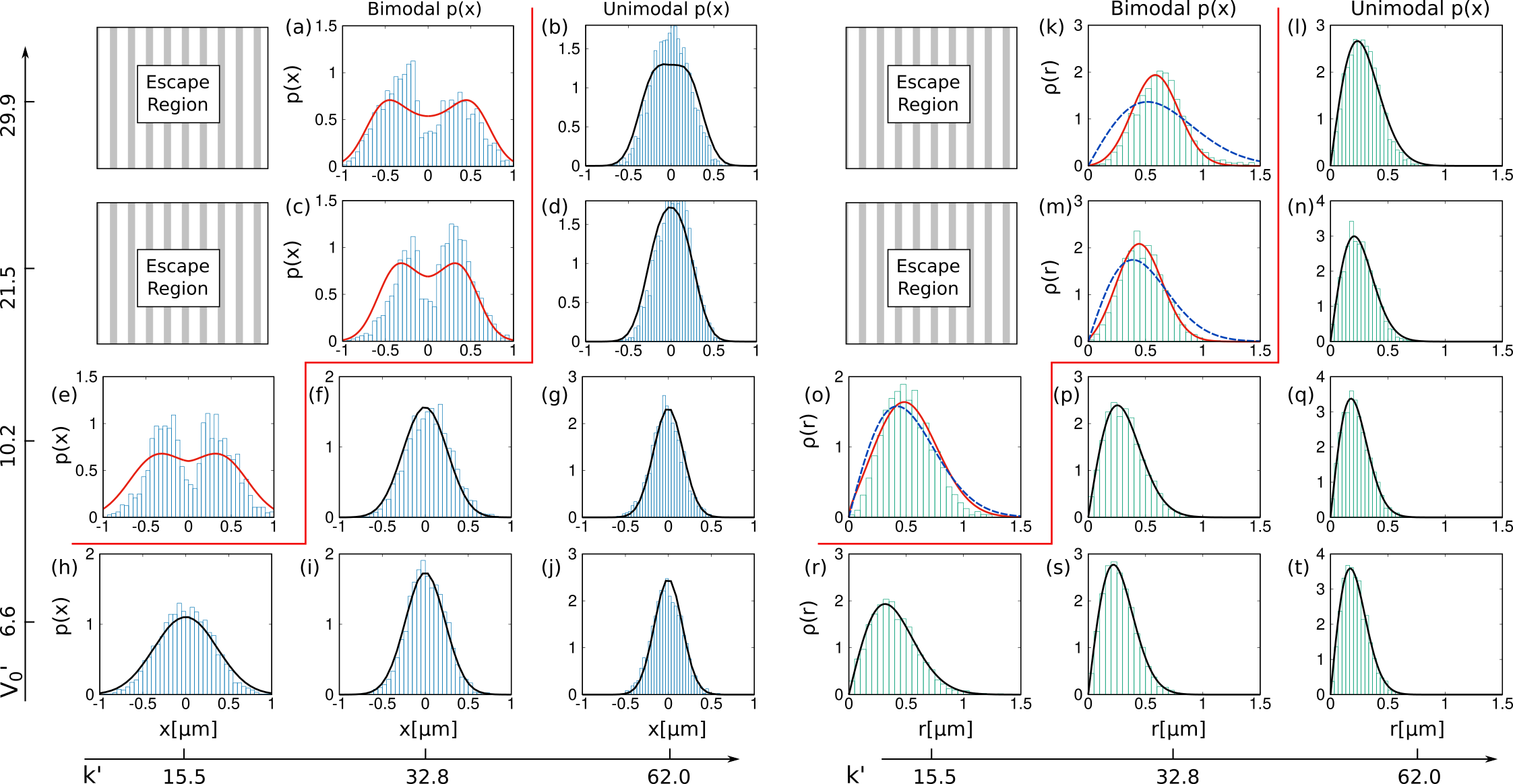}
\caption{Confined active colloids. (a-j) Normalized density distribution $\rm p(x)$ (along a single axis $\rm x$) and (k-t) radial density distribution $\rm \rho(r)$ (along the radial axis $\rm r$) for different values of reduced potential strength $k'=\tau_{\rm R} k/\gamma$ and reduced swim velocity $\rm V_0'=V_0 \sqrt{\tau_{\rm R}/{\rm D}}$. The grey panels (top, left) are drawn for the regimes of parameters where active particles escape from the trap. The two red solid lines between panels qualitatively mark the transition between bimodal and unimodal $\rm p(x)$. The histograms and solid lines are obtained from experiments (each trajectory is measured for approximately $10$ minutes) and theoretical predictions (or simulations), respectively. In particular, the red lines in panels (a), (c) and (e) are calculated from Eq.~\eqref{eq:ABP_harmonic_largepersistence}, while the red lines in panels (k), (m) and (o) are obtained by integrating Eq.~\eqref{eq:ABP_harmonic_largepersistence} along one of the two spatial coordinates. The dashed blue lines in panels (k), (m), (o) are the results of best (radial) Gaussian fits, $\rm f(r)=  b \, r\, \exp{\left(- a r^2\right)}$ where $\rm a$ and $\rm b$ are two fitting parameters. Finally, the solid black curves have been obtained through numerical simulations of the dynamics (Eq.~\eqref{eq_t1}). 
}\label{Fig3}
\end{figure*}

Confined active Brownian particles display different properties with respect to their `passive' counterparts. While the dynamics of a `passive' colloid in a harmonic trap is fully determined by the competition between thermal fluctuations and the confining force (see Eq.~\eqref{eq:passivedynamics}), active microbeads have an additional effective swimming (self-propulsion) force of magnitude $\rm F_S=6\pi\eta \rm R V_0$. If the force due to activity is smaller than the confining force, the swimmer remains confined in the harmonic potential; otherwise, the optical trap is too weak to hold back the active colloid, which will escape when $\rm F_S$ exceeds the optical force exerted on the particles at the boundary, where the optical potential reaches a maximum. This behaviour is illustrated in the Supplementary Videos S4 and S5. In S4, a colloid swimming at $\rm V_0 = 1.7$ $\rm \mu m/s$ ($\rm E=40$ $\rm V/mm$) is confined in an optical trap of stiffness $k=0.085$ $\rm pN/\mu m$. Assuming a displacement $\rm x=1$ $\rm \mu m$ with respect to the trap’s centre, the optical restoring force, ${\rm F(x)}=k{\rm x}=85$ $\rm fN$, is approximately twice the effective swimming force. The particle immediately leaves the trap if the laser is turned off (Video S4) or the self-propulsion velocity is increased \textit{in situ} to $\rm V_0 = 5$ $\rm \mu m/s$ ($\rm E=66$ $\rm V/mm$, $\rm F_S = 100$ $\rm fN$, Video S5), while keeping the stiffness of the laser trap constant.

We consider the case of fully trapped active particles and model their dynamics through a popular model in the field of active matter, known as active Brownian particles (ABP)~\cite{FilyPRL2012, ButtinoniPRL2013, SolonNatPhys2015, DigregorioPRL2018, CapriniPRL2020, tenHagen_Brownian}. The position of the colloid, $\mathbf{x}$, is described by an overdamped equation of motion in two dimensions,

\begin{equation}
	\rm \gamma\dot{\mathbf{x}}=  {\gamma}\sqrt{2 D} \textbf{W} + \mathbf{F}(\mathbf{x}) + \gamma V_0\mathbf{n},
	\label{eq_t1}
\end{equation}

\noindent which resembles Eq.~\eqref{eq:passivedynamics} except for the additional force, $\gamma {\rm V_0}\mathbf{n}$, accounting for the self-propulsion mechanism. The swim velocity, $\rm V_0$, is directed onto the orientational vector $\rm \mathbf{n}=(\cos{\theta}, \sin{\theta})$, where $\theta$ corresponds to the orientational angle of the active particle. According to the ABP model, $\theta$ evolves as
\begin{equation}
	\rm \dot{\theta} = \sqrt{\frac{2}{\tau_{\rm R}}}\xi,
	\label{eq_thetaABP}
\end{equation}

\noindent where $\xi$ is another Gaussian random variable with zero mean and unit variance, statistically independent from $\rm \textbf{W}$. We identify two dimensionless parameters that control the dynamics in the harmonic confinement: the first can be interpreted as a reduced spring constant $k'=\tau_{\rm R} k/\gamma$, given by the ratio between the rotational diffusion time and the time due to the external optical force; the second corresponds to a reduced swim velocity, $\rm V_0'=V_0 \sqrt{\tau_{\rm R}/D}$, which quantifies the strength of active and passive fluctuations (note that $\rm V_0^2 \tau_{\rm R}$ is the diffusion of potential-free active particles). In practice, $k'$ and $\rm V_0'$ are experimentally modified through $k$ and $\rm V_0$, as $\tau_{\rm R}$, $\gamma$ and $\rm D$ are constant.

Following Ref.~\cite{Caprini_Parental} (see also Sec.~4 of Ref.~\cite{Caprini2019Comparative}), we can also obtain an analytical approximation for the radial density distribution $\rm \rho(r)$, holding for $k' \gg 1$ (experimentally satisfied for all the values of $k$) and for self-propulsion velocities that are large enough (so that $\rm V_0' \gg 1$):

\begin{equation}
	\label{eq:ABP_harmonic_largepersistence}
{\rm \rho(r)\sim r^{1/2}} \exp{\left(-\frac{1}{2{\rm D}}\left(k+\frac{1}{2\tau_{\rm R}}\right) \left({\rm r}-\frac{\rm V_0}{k+(2\tau_{\rm R})^{-1}}\right)^2\right)}.
\end{equation}

\noindent The integration of Eq.~\eqref{eq:ABP_harmonic_largepersistence} over one of the two Cartesian components provides a prediction for $\rm p(x)$ (a simple explicit expression for this integral is unknown in the literature to the best of our knowledge).

In Fig.~\ref{Fig3}, we investigate the reduced distribution in Cartesian coordinates projected onto the $\rm x$-axis, $\rm p(x)$, and the radial distribution, $\rho(\rm r)$, for different values of $k'$ and $\rm V_0'$. The histograms are experimental data, whereas the solid lines correspond to numerical or theoretical predictions. When applicable, the analytical result of Eq.~\eqref{eq:ABP_harmonic_largepersistence} is used  and marked in red to distinguish it from ABP simulations. Grey panels are inserted in correspondence of the values of $k'$ and $\rm V_0'$ for which the active particles escape the harmonic trap. The shape of $\rm p(x)$ reveals a transition from a unimodal (Gaussian) to a bimodal (non-Gaussian) distribution, when the reduced swim velocity $\rm V_0'$ increases (faster particles) at fixed $k'$ (see Fig.~\ref{Fig3}~(h) $\to$ (e) and (c) $\to$ (a)), or when $k'$ is decreased (weaker traps) at fixed $\rm V_0'$ (see Fig.~\ref{Fig3}~(a) $\to$ (b); (c) $\to$ (d) and (e) $\to$ (f)). In correspondence with this transition, the peak of $\rho(\rm r)$ shifts towards a larger value of $\rm r$ compared to the Gaussian cases, and even much larger as $\rm V_0'$ is increased (Fig.~\ref{Fig3}~(k), (m), (o)). In the bimodal regime (Fig.~\ref{Fig3}~(a), (c), (e)), $\rm p(x)$ shows two symmetric peaks far from the minimum of the potential trap, where $\rm p(x)$ is depleted: an active particle persistently moves against the potential force ({\sl i.e.} it climbs up the potential barrier) until its self-propulsion is balanced by the optical force.  

Eq. \eqref{eq:ABP_harmonic_largepersistence} reflects the bimodality of the density distribution, shown in simulations of ABP confined in a harmonic trap~\cite{Malakar2020, Pototsky2012, Hennes2014, Rana2019, Basu2019, Santra2021}, and provides a distinct (and needed) feature compared to the results obtained through other popular models in active matter \cite{Szamel2014, Das2018, Dabelow2021, Caprini2021inertial, Nguyen2021}. As shown in Fig.~\ref{Fig3}~(k), (m) and (o), formula~\eqref{eq:ABP_harmonic_largepersistence} is also in good agreement with experiments: the expression for $\rm \rho(r)$ (red curve) is roughly centred around ${\rm V_0}/[k/\gamma+1/(2\tau_{\rm R})] \sim {\rm V_0}\gamma/k$ (for $k\tau_{\rm R}/\gamma \gg 1$) and accounts for the shift of the main peak observed in Fig.~\ref{Fig3}. Remarkably, the best Gaussian fits reported in the same panels (dashed blue lines) confirm the non-Gaussian effects occurring in $\rm \rho(r)$. A fair agreement between theory and experiments (less good than for $\rm \rho(r)$ because of worse statistics) is also obtained for $\rm p(x)$, where the bimodality and the positions of the peaks are correctly predicted by the integral form of Eq.~\eqref{eq:ABP_harmonic_largepersistence}. In the Gaussian cases (all the other panels in Fig.~\ref{Fig3}), experiments are reproduced through numerical simulations of the dynamics (Eq.~\eqref{eq_t1}).

\section{Discussion}

We investigated active colloids that break symmetry to achieve self-propulsion, while conserving the uniform optical properties needed to confine them in harmonic laser potentials. After verifying independently that the patchy microspheres (1) undergo free active Brownian motion in the absence of confinements and (2) are laser-tweezed just as homogenous colloids in the absence of activity, we studied the active motion in harmonic traps as a function of the curvature of the optical potential, $k$, and swimming velocity, $\rm V_0$. 

The active dynamics in the trap strongly differs from the one of `passive' Brownian colloids; the Gaussian behaviour breaks down when the swimmers are fast or when the harmonic potential is weak.  Nonetheless, a remarkable observation is that the Boltzmann-like statistics (Gaussian with effective temperature) is retained in strongly confining potential wells. At first glance, our result qualitatively disagrees with Ref.~\cite{Takatori2016Acoustic}, where Janus particles (platinum/polystyrene), swimming in hydrogen peroxide solution and harmonically confined in acoustic traps, show non-Boltzmann distributions for stiff traps (large values of $k$) and Gaussian distributions for weak traps (small values of $k$). In Ref.~\cite{Takatori2016Acoustic}, the transition is induced by the increase of the reduced spring constant, $k'$ (in particular, when $k'$ crosses one), interpreted in terms of the competition between persistence length, $\rm V_0 \tau_{\rm R}$, and the effective width of the potential, ${\rm V_0} k/\gamma$. On the contrary, our study focuses on the regime $k' \gg 1$ for all the values of the trap stiffness considered. In this regime, the transition (Gaussian $\to$ non-Gaussian shape of $\rm p(x)$) is induced by the increase of the reduced swim velocity, $\rm V_0'$, and the decrease of the potential stiffness, $k'$. This occurs because stronger traps reduce the ability of active particles to climb up the potential barrier so that they are only responsible for a reduction of the variance of $\rm p(x)$ without affecting its Gaussian shape (see Fig.~\ref{Fig3}~(g) $\to$ (f); (j) $\to$ (i) $\to$ (h)). This effect is qualitatively in agreement with our intuition coming from the passive scenario. However, unlike the case of passive particles, the variance of $\rm p(x)$ does not scale as $\sim 1/k$ (not shown). This observation implies the occurrence of non-Boltzmann distributions also in the regime of parameters where $\rm p(x)$ is characterised by a Gaussian shape. Similarly, the increase of $\rm V_0'$ (for low $\rm V_0'$ and large values of $k'$) leads to the variance increase of the distribution, showing a phenomenology qualitatively consistent with the increase of the effective active temperature (see Fig.~\ref{Fig3}~(i) $\to$ (f); (j) $\to$ (g) $\to$ (d)). Moreover, its dependence on the model parameters goes beyond the simple scaling $\sim \rm V_0^2/D_R$ (not shown), {\sl i.e.} the effective diffusion typical of an unconfined active particle. The breaking of these scalings of the variance ($\sim 1/k$ and $\sim \rm V_0^2 \tau_R$) occurs because the system is far from the equilibrium for all the combinations of parameters of $k'$ and $\rm V_0'$ experimentally investigated.

Finally, we highlight the advantages of our experimental system. Firstly, the confining and swimming force ($ {\rm \mathbf{F}}=-k{\rm \mathbf x}$ and $\rm \mathbf{F}_{S}= \gamma {\rm V_0}\mathbf{n}$) have similar magnitude and can be both tuned \textit{in situ} -- by changing the laser power and the strength of the AC electric field, respectively -- to optically trap or release \textit{on demand} the microswimmers. Secondly, the optical traps have a size comparable to the particle's diameter, as opposed to existing experimental works of active colloids in harmonic wells \cite{Takatori2016Acoustic,Dauchot_Dynamics}, where the confinements are much larger. Lastly, unlike many experiments in which active particles are used in combination with laser fields \cite{Pince_Disorder,Lozano_Phototaxis,Lavergne_Group}, here the optical beam does not alter the self-propulsion mechanism, but only provides a confining potential. All in all, these features are paramount to confine two particles in distinct traps and measure the pair interaction as a function of the relative distance, or study active Brownian motion in complex potential landscapes (e.g. periodic or random substrate potentials, time-modulated traps etc.).

\acknowledgments
We thank Stefan Egelhaaf for access to his optical-tweezing setup, which has been used for some experiments presented in this manuscript. LC acknowledges support from the Alexander Von Humboldt foundation. LA  acknowledges financial support from the Swiss National Foundation (SNF) Spark grant, No 172913, and is grateful to Jose-Luis Pujol for assistance with the deposition of silicon dioxide.

\bibliographystyle{rsc}

%\bibliography{bib.bib}

\begin{thebibliography}{0}


\bibitem{Grier_Review}
  \Name{Grier D. G.}
  \REVIEW{Curr. Opin. Colloid Interface Sci}{2}{1997}{264-270}
  
\bibitem{Optical Tweezers}
  \Name{Jones P. H., Marag{\'o} O. M. \and Volpe G.}
  \Book{Optical tweezers: principles and applications}
  \Publ{Cambridge University Press, Cambridge UK}
  \Year{2015}

\bibitem{Volpe_Review}
  \Name{Gieseler J., Gomez-Solano J. R., Magazz{\'u} A., Castillo I. P., Garcia L. P., Gironella-Torrent M.,   Viader-Godoy X., Ritort F., Pesce G., Arzola A. V., Volke-Sepulveda K. \and Volpe G.}
  \REVIEW{Adv. Opt. Photonics}{13}{2021}{74-241}.
  
\bibitem{Crocker_Entropic}
  \Name{Crocker J. C., Matteo J. A., Dinsmore A. D. \and Yodh A. G.}
  \REVIEW{Phys. Rev. Lett.}{82}{1999}{4352}    
  
\bibitem{Buttinoni_Mechanical}
  \Name{Buttinoni I. \and Dullens R. P. A.}
  \REVIEW{J. Phys. Mater.}{4}{2021}{025001}  
  
\bibitem{Park_Heterogeous}
  \Name{Park B. J., Vermant J. \and Furst E. M.}
  \REVIEW{Soft Matter}{46}{2010}{5327-5333} 
  
\bibitem{Mittal_Polarization}
 \Name{Mittal M., Lele P. P., Kaler E. W. \and Furst E. M.}
 \REVIEW{J. Phys. Chem.}{129}{2008}{064513} 
   
\bibitem{ElMasri_Measuring}
  \Name{El Masri D., van Oostrum P., Smallenburg F., Vissers T., Imhof A., Dijkstra M. \and van Blaaderen A.}
  \REVIEW{Soft Matter}{7}{2011}{3462-3466}  
  
\bibitem{Fadanelli_Measurement}
  \Name{Carrasco-Fadanelli V. \and Castillo R.}
  \REVIEW{Soft Matter}{15}{2019}{5815-5818}   
 
\bibitem{Franosch_Resonance}
 \Name{Franosch T., Grimm M., Beluschkin M., Mor F. M., Foffi G., Forr{\'o} M. \and Jeney S.}
 \REVIEW{Nature}{478}{2011}{85-88}    
 
\bibitem{Williams_Direct}
 \Name{Williams I., O\u{g}uz E. C., Bartlett P., L\"{o}wen H. \and Royall C. P.}
 \REVIEW{Nat. Commun.}{4}{2013}{1-6}   
  
\bibitem{Gammaitoni_Stochastic}
 \Name{Gammaitoni L., H\"{a}nggi P., Jung P. \and Marchesoni F.}
 \REVIEW{Rev. Mod. Phys.}{70}{1998}{223}  
  
\bibitem{Simon_Escape}
 \Name{Simon A.\and Libchaber A.}
 \REVIEW{Rev. Mod. Phys.}{66}{1992}{3375}  

\bibitem{Ciliberto_Experiments}
 \Name{Ciliberto S.}
 \REVIEW{Phys. Rev. X}{7}{2017}{021051}  
    
\bibitem{Bickle_Realization}
 \Name{Bickle V. \and Bechinger C.}
 \REVIEW{Nat. Phys.}{8}{2012}{143-146}  
  
\bibitem{Martinez_Brownian}
 \Name{Martinez I. A., Rold{\'a}n E., Dinis L., Petrov D., Parrondo J. \and Rica R. A.}
 \REVIEW{Nat. Phys.}{12}{2016}{67-70}  
  
\bibitem{Admon_Experimental}
 \Name{Admon T., Rahav S. \and Roichman Y.}
 \REVIEW{Phys. Rev. Lett.}{121}{2018}{180601}  

\bibitem{Simpson_Inhomogeneous}
 \Name{Simpson S. H.}
 \REVIEW{J. Quant. Spectrosc. Radiat. Transf.}{146}{2014}{81-99} 

\bibitem{Purcell}
\Name{Purcell E. M.}
\REVIEW{Am. J. Phys.}{45(1)}{1977}{3-11}.

\bibitem{Bechinger_Active}
\Name{Bechinger C., Di Leonardo R., L\"{o}wen H., Reichhardt C., Volpe G. \and Volpe G.}
\REVIEW{Rev. Mod. Phys.}{88}{2016}{045006}

\bibitem{Aubret_Eppur}
 \Name{Aubret A., Ramananarivo S. \and Palacci J.}
 \REVIEW{Curr. Opin. Colloid Interface Sci}{30}{2017}{81-89} 

\bibitem{Aubret_Metamachines}
 \Name{Aubret A., Martinet Q. \and Palacci J.}
\REVIEW{Nat. Commun.}{12}{2021}{1-9}

\bibitem{Lawson_Optical}
 \Name{Lawson J. L., Jenness N. J. \and Clark R. L.}
 \REVIEW{Opt. Express}{23}{2015}{33956-33969}

\bibitem{Zong_Optically}
 \Name{Zong Y., Liu J., Liu R, Guo H., Yang M., Li Z. \and Chen K.}
 \REVIEW{ACS Nano}{9}{2015}{10844-10851}
 
\bibitem{Moyses_Trochoidal}
 \Name{Moyses H., Palacci J., Sacanna S. \and Grier D. G.}
 \REVIEW{Soft Matter}{12}{2016}{6357-6364} 
 
\bibitem{Pince_Disorder}
 \Name{Pin\c{c}e E., Velu S. K., Callegari A., Elahi P., Gigan S., Volpe G. \and Volpe G.}
 \REVIEW{Nat. Commun.}{7}{2016}{1-8}
 
\bibitem{Lozano_Phototaxis}
 \Name{Lozano C., Ten Hagen B., L\"{o}wen H. \and Bechinger C.}
 \REVIEW{Nat. Commun.}{7}{2016}{1-10}
 
\bibitem{Lavergne_Group}
 \Name{Lavergne F. A., Wendehenne H., B\"{a}uerle T. \and Bechinger C.}
 \REVIEW{Science}{364}{2019}{70-74}
 
\bibitem{Jahanshahi_Realization}
 \Name{Jahanshahi S., Lozano C., Liebchen B., L\"{o}wen H. \and Bechinger C.} 
 \REVIEW{Commun. Phys.}{3}{2020}{1-11}
 
\bibitem{Schmidt_Non-equilibrium}
 \Name{Schmidt F., \u{S}ipov\'{a}-Jungov\'{a} H., K\"{a}ll M., W\"{u}rger A. \and Volpe G.} 
 \REVIEW{Nat Commun.}{12}{2021}{1-9}
 
\bibitem{Caprini_Parental}
 \Name{Caprini L., Sprenger A. R., L\"{o}wen H. \and Wittmann R.}
 \REVIEW{J. Chem. Phys.}{156}{2022}{071102} 
 
\bibitem{Pototsky2012}
 \Name{Pototsky A. \and Stark H.}
 \REVIEW{Europhys. Lett.}{98}{2012}{50004} 

\bibitem{Szamel2014}
 \Name{Szamel G.}
 \REVIEW{Phys. Rev. E}{90}{2014}{012111} 
 
\bibitem{Jahanshahi_Brownian}
 \Name{Jahanshahi S., L\"{o}wen H. \and Ten Hagen B.}
 \REVIEW{Phys. Rev. E}{95}{2017}{022606} 
 
\bibitem{Chaudhuri_Active} 
 \Name{Chaudhuri D. \and Dhar A.}
 \REVIEW{J. Stat. Mech.}{1}{2021}{013207} 
 
\bibitem{Khadem_Delayed} 
 \Name{Khadem S. M. J. \and Klapp S. H.}
 \REVIEW{Phys. Chem. Chem. Phys.}{21}{2019}{13776-13787} 

\bibitem{Dauchot_Dynamics}
 \Name{Dauchot O. \and D\'{e}mery V.}
 \REVIEW{Phys. Rev. Lett.}{122}{2019}{068002} 

\bibitem{Takatori2016Acoustic}
 \Name{Takatori S., De Dier R., Vermant J. \and Brady J. F.}
 \REVIEW{Nat. Commun.}{7}{2016}{10694} 
 
 \bibitem{Rodriguez_Self}
 \Name{Rodr\'{i}guez-de Marcos L. V., Larruquert J. I., M\'{e}ndez J. A. \and Azn\'{a}rez J. A.} 
 \REVIEW{Opt. Mater. Express}{6}{2016}{3622-3637}
 
 \bibitem{Gao_Exploitation}
 \Name{Gao L., Lemarchand F. \and Lequime M.} 
 \REVIEW{Opt. Express}{20}{2012}{15734-15751} 

\bibitem{Ristenpart_Electrohydrodynamic}
 \Name{Ristenpart W. D., Aksai I. A. \and Saville D. A.}
 \REVIEW{J. Fluid. Mech.}{575}{2007}{83-109} 

\bibitem{Ma_Inducing}
 \Name{Ma F., Yang X., Zhao H. \and Wu N.}
 \REVIEW{Phys. Rev. Lett.}{115}{2015}{208302} 

\bibitem{Ni_Hybrid}
 \Name{Ni S., Marini E., Buttinoni I., Wolf H. \and Isa L.}
 \REVIEW{Soft Matter}{13}{2017}{4252-4259} 

\bibitem{Alvarez_Reconfigurable}
 \Name{Alvarez L., Fernandez-Rodriguez M. A., Alegria A., Arrese-Igor S., Zhao K., Kr\"{o}ger M. \and Isa L.}
 \REVIEW{Nat. Commun.}{12}{2021}{1-9} 

\bibitem{Fernandez_Feedback}
 \Name{Fernandez-Rodriguez M. A., Grillo F., Alvarez L., Rathlef M., Buttinoni I., Volpe G. \and Isa L.}
 \REVIEW{Nat. Commun.}{11}{2020}{1-10} 

\bibitem{Gangwal_Induced}
 \Name{Gangwal S., Cayre O. J., Bazant M. Z. \and Velev O. D.}
 \REVIEW{Phys. Rev. Lett.}{100}{2008}{058302} 
 

 
 





\bibitem{FilyPRL2012}
 \Name{Fily Y. \and Marchetti M. C.}
 \REVIEW{Phys. Rev. Lett.}{108}{2012}{235702} 

\bibitem{ButtinoniPRL2013}
 \Name{Buttinoni I., Bialk\'{e} J., K\"{u}mmel F., L\"{o}wen H., Bechinger C. \and Speck T.}
 \REVIEW{Phys. Rev. Lett.}{110}{2013}{238301} 


\bibitem{SolonNatPhys2015}
 \Name{Solon A. P., Fily Y., Baskaran A., Cates M. E., Kafri Y., Kardar M. \and Tailleur J.}
 \REVIEW{Nat. Phys.}{11}{2015}{673-678} 

\bibitem{DigregorioPRL2018}
 \Name{Digregorio P., Levis D., Suma A., Cugliandolo L. F., Gonnella G. \and Pagonabarraga I.}
 \REVIEW{Phys. Rev. Lett.}{121}{2018}{098003} 
 
\bibitem{tenHagen_Brownian}
\Name{ten Hagen B., van Teeffelen S. \and L\"{o}wen H.}
\REVIEW{J. Phys. Condens. Matter}{23}{2011}{194119} 




%inserted
\bibitem{CapriniPRL2020}
 \Name{Caprini L., Marini Bettolo Marconi U. \and Puglisi A.}
 \REVIEW{Phys. Rev. Lett.}{124}{2020}{078001} 
%Caprini, Lorenzo, U. Marini Bettolo Marconi, and Andrea Puglisi. "Spontaneous velocity alignment in motility-induced phase separation." Physical review letters 124.7 (2020): 078001.

% reference for the prediction
%inserted
\bibitem{Caprini2019Comparative}
 \Name{Caprini L., Hernández-García E., López C. \and Marini Bettolo Marconi U.}
 \REVIEW{Sci. Rep.}{9}{2019}{16687} 

%%REFERENCES FOR ABP in harmonic trap: theory and simulations
%%%%%%%%%%%%%%

%inserted
\bibitem{Malakar2020}
 \Name{Malakar K., Das A., Kundu A. , Kumar K. V. \and Dhar A. }
 \REVIEW{Phys. Rev. E}{101}{2020}{022610} 
%K. Malakar, A. Das, A. Kundu, K. V. Kumar, and A. Dhar, Phys. Rev. E 101, 022610 (2020). https://doi.org/10.1103/physreve.101.022610

%inserted


%inserted
\bibitem{Hennes2014}
\Name{Hennes M., Wolff K. \and Stark H.}
\REVIEW{Phys. Rev. Lett.}{112}{2014}{238104} 
%Hennes, M., Wolff, K. \and Stark, H. Phys. Rev. Lett. 112, 238104 (2014). https://doi.org/10.1103/physrevlett.112.23810476. 

%inserted
\bibitem{Rana2019}
\Name{Rana S., Samsuzzaman M. \and Saha A.}
\REVIEW{Soft Matter}{15}{2019}{8865} 
%S. Rana, M. Samsuzzaman, and A. Saha, Soft Matter 15, 8865 (2019). https://doi.org/10.1039/c9sm01691k77. 

%inserted 
\bibitem{Basu2019}
\Name{Basu, U. Majumdar, S. N. Rosso, A. \and Schehr G.}
\REVIEW{Phys. Rev. E}{100}{2019}{062116} 
%U. Basu, S. N. Majumdar, A. Rosso, and G. Schehr, Phys. Rev. E 100, 062116 (2019). https://doi.org/10.1103/physreve.100.06211678. 

%inserted
\bibitem{Santra2021}
\Name{Santra I., Basu U. \and Sabhapandit S.}
\REVIEW{Soft Matter}{17}{2021}{10108} 
%I. Santra, U. Basu, and S. Sabhapandit, Soft Matter 17, 10108 (2021). https://doi.org/10.1039/d1sm01118a
 

%%%%%%%%%%%%% REFERENCES FOR other models: AOUP UNDER HARMONIC CONFINEMENT
%
%inserted


%inserted
\bibitem{Das2018}
\Name{Das S., Gompper G. \and Winkler R. G.}
\REVIEW{New J. Phys.}{20}{2018}{015001} 
%S. Das, G. Gompper, and R. G. Winkler, New J. Phys. 20, 015001 (2018). https://doi.org/10.1088/1367-2630/aa9d4b69. 
 
%inserted
\bibitem{Dabelow2021}
\Name{Dabelow L., Bo S. \and Eichhorn R.}
\REVIEW{J. Stat. Mech.}{2021}{2021}{033216} 
%L. Dabelow, S. Bo, and R. Eichhorn, J. Stat. Mech.: Theory Exp. 2021, 033216. https://doi.org/10.1088/1742-5468/abe6fd

%inserted
\bibitem{Caprini2021inertial}
\Name{Caprini L. \and Marini Bettolo Marconi U.}
\REVIEW{J. Chem. Phys.}{154}{2021}{024902} 
%L Caprini, U Marini Bettolo Marconi, The Journal of Chemical Physics 154 (2), 024902, 2021

%inserted
\bibitem{Nguyen2021}
\Name{Nguyen G.H.P., Wittmann R. \and Löwen H.}
\REVIEW{J. Phys. Cond. Matt.}{34}{2021}{035101} 
%GHP Nguyen, R Wittmann, H Löwen Journal of Physics: Condensed Matter 34 (3), 035101, 2021



\end{thebibliography}

\end{document}